\documentclass[fleqn,twoside]{article}
\usepackage{espcrc2}

\usepackage{graphics}
\usepackage[figuresright]{rotating}

\newcommand{\AmS}{{\protect\the\textfont2
  A\kern-.1667em\lower.5ex\hbox{M}\kern-.125emS}}

\title{A Novel Two-Step Laser Ranging Technique for a Precision Test of the
Theory of Gravity}

\author{Konstantin Penanen and Talso Chui \address[JPL]{Jet Propulsion
Laboratory, California Institute of Technology, \\
4800 Oak Grove Drive, Pasadena, CA 91109, USA}
\thanks{We would like to thank S.G. Turyshev and M.M. Nieto for helpful discussions. This work was carried out at the Jet Propulsion Laboratory, California
        Institute of Technology under a contract with the National Aeronautics and
        Space Administration. Contact: T. C. P. Chui (818)354-3104,
        Talso.C.Chui@jpl.nasa.gov}}

\begin{document}

\begin{abstract}
All powered spacecraft experience residual systematic acceleration
due to anisotropy of the thermal radiation pressure and fuel
leakage. The residual acceleration limits the accuracy of any test
of gravity that relies on the precise determination of the
spacecraft trajectory. We describe a novel two-step laser ranging
technique, which largely eliminates the effects of non-gravity
acceleration sources and enables celestial mechanics checks with
unprecedented precision.  A passive proof mass is released from
the mother spacecraft on a solar system exploration mission.
Retro-reflectors attached to the proof mass allow its relative
position to the spacecraft to be determined using optical ranging
techniques.  Meanwhile, the position of the spacecraft relative to
the Earth is determined by ranging with a laser transponder.  The
vector sum of the two is the position, relative to the Earth, of
the proof mass, the measurement of which is not affected by the
residual accelerations of the mother spacecraft.  We also describe
the mission concept of the Dark Matter Explorers (DMX), which will
demonstrate this technology and will use it to test the hypothesis
that dark matter congregates around the sun.  This hypothesis
implies a small apparent deviation from the inverse square law of
gravity, which can be detected by a sensitive experiment.  We
expect to achieve an acceleration resolution of $\sim 10^{-14}
m/s^2$. DMX will also be sensitive to acceleration towards the
galactic center, which has a value of $\sim 10^{-10} m/s^2$. Since
dark matter dominates the galactic acceleration, DMX can also test
whether dark matter obeys the equivalence principle to a level of
100 ppm by ranging to several proof masses of different
composition from the mother spacecraft. \vspace{1pc}
\end{abstract}

\maketitle

\section{Introduction}

Precise observations of motion of celestial bodies, along with
corresponding theoretical models have formed the foundation of
modern science. These include Kepler's insight into planetary
motion based largely on Mars' ephemeris, Newton's subsequent
formulation of the universal law of gravitation, and Einstein's
explanation of Mercury perihelion advance. With increasing
precision of modern measurement techniques it has become possible
to determine the range to objects on the Moon to a centimeter
level. Similar precision is routinely achieved in artificial
satellite ranging. Science return of such measurements include
tests of strong equivalence principle (Lunar Laser Ranging (LLR),
\cite {EP1,EP2}) and general relativity tests (LAGEOS series
\cite{lageos}). While the precision of the range determination
improves, the effects of non-gravitational influences on the
studied objects becomes more significant. The observed anomalous
$\sim 9\times 10^{-10} m/s^2$ acceleration of the Pioneer
spacecraft towards the Sun \cite{Pioneer} is an example of a
potentially significant discovery obscured by the possible
non-gravitation systematic effects. Pressure from $\sim 60$ Watts
of unisotropically radiated heat from the spacecraft may account
for the entire effect. Other typical non-gravitational influences
include solar radiation and wind, propellant leakage, pressure due
to directed radio communication, etc. Several techniques have been
used to avoid or to mitigate the effects of non-gravitational
systematics. Such effects would be small for objects that have
large mass to surface area ratios. Laser ranging to
retro-reflectors left by the Appollo and the Luna missions on the
surface of the Moon, and the radio ranging to the Viking
spacecraft on Mars make use of large masses of celestial bodies.
Similar approach can be used by powered spacecraft only on or near
a planet or an asteroid. LAGEOS Earth-orbiting retro-reflector
satellites reduce the thermal radiation effects by not carrying
power sources. Another ingenious work-around is the use of
drag-free satellites (DISCOS, TRIAD, TIP, NOVA, LISA, STEP) where
an inner object is shielded from the non-gravitational forces by
an outer shell, which actively follows it. This technique requires
substantial propellant expenditure. Moreover, in order to avoid
gravitational interaction of the inner and outer bodies, the mass
distribution of the spacecraft has to be extraordinarily
well-defined.

In the context of gravitational science return from trajectory
determination, the prospect of a nuclear-powered spacecraft with
associated high-power communication subsystem presents a dilemma.
With technology currently employed in lunar and satellite laser
ranging, an optical transponder onboard would allow distance to
such spacecraft to be measured with centimeter precision over
hundreds of AU. At the same time, anisotropy in thermal radiation
pressure due to the power source would alter the trajectory,
making such a measurement meaningless. In this paper, we propose a
novel two-step laser ranging technique which will circumvent the
non-gravitational systematic acceleration and will allow
centimeter precision ephemeris determination on the 100-1000 AU
scale.

\section{Tracking configuration}
\begin{figure}
\scalebox{.75}{\includegraphics{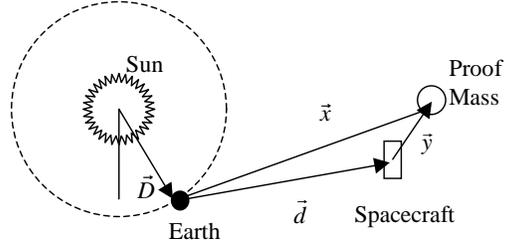}} \caption{Illustration of
the two-step ranging technique. Spacecraft experiences
unpredictable parasitic acceleration, with resultant error in
$\vec{d}$ and $\vec{y}$. Position of the proof mass relative to
Earth, $\vec{x}$, is unaffected by spacecraft motion and can be
determined from simultaneous measurement of $\vec{d}$ and
$\vec{y}$, $\vec{x}=\vec{d}+\vec{y}$. Range from Earth to the
proof mass, $\|\vec{x}\|$ can be determined to a few $cm$.}
\end{figure}
 Figure 1 illustrates the concept of the two-step ranging
technique. Several proof masses, ~1kg each, of different
compositions will be released from a mother spacecraft.
Retro-reflectors will be attached to the surface of the proof
masses.  The mother spacecraft will follow the proof masses at a
distance of $\sim 1 km$ away, so that the gravitational and
electrostatic forces from the spacecraft would not affect the
proof masses significantly. The distances to the proof masses will
be determined by laser ranging from the mother spacecraft. The
relative orientation of the proof masses will be determined by an
optical telescope. The relative position of the proof masses can
be determined with centimeter precision in a single measurement.
The range to the spacecraft relative to the Earth will be inferred
from the time it takes a laser beam to travel from the Earth to
the spacecraft and back. The measurement is done using a laser
transponder onboard the spacecraft, which sends back a reply beam
to Earth after it detects a beam from Earth.  Any time delay will
be measured, and the information will be sent back to Earth
together with the positions of the proof masses at a later time.
The range of the proof mass relative to the Earth can then be
determined. While the range to the Earth and and relative position
of the proof masses are affected by residual motion of the mother
spacecraft due to fuel leakage and uncontrolled thermal radiation
pressure, the vector sum is not.

\section{Imaging and ranging} The ranging laser on the spacecraft
will be similar to the high-power YAG pulse lasers used in LLR
\cite{Dickey}. The wavelength is  $\sim 1\mu m$, the beam
divergence from a $25-cm$ telescope is  $\sim 10^{-5}$ radian, the
pulse width is $\sim 100ps$, the energy for each pulse is $\sim
100 mJ$, a few pulses are fired in a train, and the pulse train is
repeated at a rate of $10 Hz$. Thus the mean beam power for the
laser is $P\approx 10W$. With a 1-meter telescope on Earth or on a
near-Earth orbit and assuming quantum efficiency of $0.2$, we
estimate $\sim 4$ photons per second detected by the receiver when
the spacecraft is $1000 AU$ away. This rate is higher than that in
a typical LLR experiment, where a reflected photon is detected
every few seconds \cite{EP2}. If the error in range determination
is limited by the $100ps$ pulse width, the range to the spacecraft
can be determined to $\sim 3cm$. Further improvement can be
achieved by analyzing the pulse shape. Pulse coding and time
shutter techniques similar to those of LLR can also be used. A
similar on-board telescope in conjunction with a CCD will be used
to image the proof masses. Estimated diffraction-limited lateral
resolution at $1km$ distance is $\sim 1cm$. Ranging from the
mother spacecraft to the proof masses will be achieved by sending
a wide-beam pulse and analyzing time delay of the reflected
signals.

\section{Error Sources}

We have evaluated known sources contributing to the error in range
determination, and potential residual sources contributing to
proof mass acceleration. Pulse width, clock drift, and atmospheric
dispersion limit the range accuracy to a few centimeters in a
30-day measurement. Solar radiation uncertainty, solar wind,
thermal radiation pressure, uncertainty in gravitational and
electrostatic interaction with the mother spacecraft set a limit
of $\sim 10^{-14} m/s^2$. In combination, the systematic
non-gravitational sources of error and errors of the range
determination will allow acceleration measurements to $10^{-14}
m/s^2$ in a measurement lasting several tens of days.

\section{Science Goals}

We propose the described ranging scheme to be implemented on a
deep space exploration mission as a guest experiment. Any
deviation from the expected gravitational interaction of the proof
masses with the solar system bodies will appear as anomalous
correction to their trajectory. This experiment can be used to
probe predictions of the Modified Newtonian dynamics and
tensor-scalar theories of gravitation on the Solar system scale,
and to search for gravitational evidence of dark matter. While not
sensitive to the average galactic dark matter density (which would
require $\sim 10^{-16} m/s^2$ acceleration resolution), the
proposed scheme should detect any dark matter concentrated by
Solar gravity by a factor of $\sim 100$ or more. Proof masses of
different isotopic composition may also detect any equivalence
principle violation in their attraction towards the galactic
center to 100ppm. The attraction is mostly due to dark matter.
Kuiper belt and Oort cloud objects may produce additional
short-term corrections to the acceleration. Quantifying
interaction with such objects will allow better understanding of
their abundance, mass and trajectory distributions.

\section{Conclusions}

At present there is a strong impetus from NASA to develop thermal
nuclear energy and optical communication capability for deep space
missions. With the proposed ranging scheme, these capabilities and
the relevant infrastructure can be used to probe gravitational
interactions on the Solar system scale with unprecedented
precision. Our analysis indicates that centimeter-level ranging
resolution, and $10^{-14} m/s^2$ acceleration resolution in a
30-day measurement can be achieved.

\end{document}